\documentclass[twocolumn,showpacs,aps,prb,preprintnumbers,amsmath,amssymb,floatfix]{revtex4}

\usepackage{graphicx}
\usepackage{dcolumn}
\usepackage{bm}
\usepackage{epsfig}

\begin{document}
\date{\today}

\title{ 
Diffusion on edges of insulating graphene with intravalley and intervalley scattering }

\author{ G. Tkachov$^1$ and Martina Hentschel$^2$}

\affiliation{ 
$^1$ Institute for Theoretical Physics and Astrophysics, University of W\"urzburg, Am Hubland, 97074 W\"urzburg, Germany,\\ 
$^2$ Technische Universit\"at Ilmenau, Institut f\"ur Physik, Weimarer Str. 25, D-98693 Ilmenau}

\begin{abstract}
Band gap engineering in graphene may open the routes towards transistor devices
in which electric current can be switched off and on at will. 
One may, however, ask if a semiconducting band gap alone is sufficient to quench the current in graphene.
In this paper we demonstrate that despite a bulk band gap graphene can still have metallic conductance along the sample edges
(provided that they are shorter than the localization length).  
We find this for single-layer graphene with a zigzag-type boundary which hosts gapless propagating edge states even in the presence  
of a bulk band gap. By generating inter-valley scattering, sample disorder reduces the edge conductance. However, for weak scattering 
a metallic regime emerges with the diffusive conductance ${\cal G} = (e^2/h) (\ell_{_{KK^\prime} } / L)$ per spin, 
where $\ell_{_{KK^\prime} }$ is the transport mean-free path due to the inter-valley scattering and $L \gg \ell_{_{KK^\prime} }$ is the edge length. 
We also take intra-valley scattering by smooth disorder (e.g. by remote ionized impurities in the substrate) into account. 
Albeit contributing to the elastic quasiparticle life-time, the intra-valley scattering has no effect on ${\cal G}$.
\end{abstract}
\maketitle

\section{Introduction}
\label{Intro}

Unique electronic properties of single-atomic-layer graphene \cite{Neto09} stem from
its two-dimensional (2D) semi-metallic energy spectrum with gapless conical conduction 
and valence bands. Engineering of a semiconductor-type band gap is expected to 
provide another desirable means of controlling electric current in graphene, 
laying the basis for electronic applications \cite{Novoselov07}. 
Several mechanisms of the gap generation in the single-layer graphene have been discussed in literature 
\cite{Novoselov07,Zhou07,BN_Gio07,Strain_Moh09,Strain_Pereira09,Strain_Guinea10}, 
including breaking the sublattice symmetry on a hexagonal boron-nitride substrate (see e.g. Ref. \onlinecite{BN_Gio07}) and 
using mechanical strain (see e.g. Refs. \onlinecite{Strain_Moh09,Strain_Pereira09,Strain_Guinea10}).  

Although the opening of the band gap could have a desirable effect on transport in the 2D bulk of the material, 
it is, generally, not sufficient to control electric conduction near sample boundaries.
Boundaries of graphene are natural extended defects that can host unusual electronic states such as edge states appearing on a zigzag boundary \cite{Fujita96}. 
Various manifestations of such edge states in transport properties and spectroscopy of graphene have been discussed in recent years 
(e.g. Refs. \onlinecite{Waka00,Peres06,Brey06,Sasaki06,Koba05,Niimi06,GT07,Akhmerov08,Burset08,Gusynin08,Yazyev08,Evaldson08,Wimmer08,Mucciolo09,Basko09,GT09a,Burset09,GT09b,Girit09,GT09c,Volkov09,Viana09,Ratnikov10,Herrera10,Qiao11,Rozhkov11,Roeder11,Gunlycke12}).
Most essential for our present discussion is the finding that the edge states remain conducting even when the 2D bulk turns into a band insulator, 
e.g. due to a staggered potential breaking the sublattice symmetry (see e.g. Refs. \onlinecite{Volkov09,Qiao11}). 
This is indeed expected since the edge states reside on one of the graphene sublattices, and, therefore, the influence of the staggered potential 
is reduced to an energy shift without dramatic changes in the edge-state dispersion. Thus, the edge states present a potential obstacle for the realization 
of the band insulator regime in graphene, providing pathways for leakage current. 
It is of both theoretical and practical interest to identify the factors that may help to reduce the edge conduction. 
As one of such factors, in this paper we theoretically consider structural disorder involving both smooth potential fluctuations, 
which couple states within the same graphene valley (intra-valley scattering), and atomically sharp defects generating inter-valley scattering.  

Earlier, the influence of disorder on the edge transport was studied numerically for the conventional semi-metallic state of graphene 
(see e.g. Refs. \onlinecite{Waka00,Wimmer08,Evaldson08,Mucciolo09,Waka09}). Interestingly, the edge transport cannot be quenched by usual potential disorder, 
e.g. by smooth potential fluctuations due to remote ionized impurities. Only atomically sharp defects suppress the edge transport 
by mixing counter-propagating edge channels through inter-valley scattering. 
In this paper we extend these findings to the edge transport in the insulating graphene 
with the staggered potential, $\Delta$. Instead of using numerical approaches, we perform analytic diagrammatic calculations of the edge conductance, 
explicitly proving that the edge-transport mean-free path $\ell_{KK^\prime}$ is limited only by scattering between graphene's two valleys, $K$ and $K^\prime$, and 
unaffected by smooth potential disorder. Our calculations indicate that the undoped insulating graphene can host diffusive metallic edge states 
with the conductance given per spin by  
\begin{eqnarray}
{\cal G} = \frac{e^2}{h} \frac{ \ell_{_{KK^\prime} } }{L}, \quad \ell_{_{KK^\prime} } =  \frac{ \hbar^2 v^2 }{ 4 w(k_+ - k_-) }, \quad k_\pm = \pm \frac{\Delta}{\hbar v}, 
\label{Cond}
\end{eqnarray}
where the transport mean-free path $\ell_{KK^\prime}$ is expressed in terms of the Fourier transform of the intervalley-disorder correlation function, $w(k)$,  
and the edge-state velocity $v$, and $k_\pm$ are the edge-state Fermi points relative to the $K$ and $K^\prime$ valleys, respectively 
[$L$ is the distance between the source and drain, assumed much larger than $\ell_{KK^\prime}$, 
but smaller than the typical size of the system that exhibits full localization \cite{Waka09}]. 
The specifics of the insulating graphene lies in the fact that the Fermi points $k_\pm$ are shifted with respect 
to $K$ and $K^\prime$ points by the staggered potential $\Delta$. Given furthermore the low edge-state velocity $v$, \cite{Peres06,Sasaki06,Akhmerov08} 
a metallic diffusive regime under weak scattering condition $|k_\pm|\ell_{_{KK^\prime} }\gg 1$ emerges without any doping of the material.
This is in stark contrast with the conventional metallic transport which occurs when the Fermi level is pushed into a conduction or valence band.  

The subsequent sections give a complete account of our theoretical approach: 
In Sec. \ref{Edge} we introduce the model for the edge states in disorder-free insulating graphene with the staggered potential and 
calculate the edge-state Green's functions.  
In Sec. \ref{Conductance} we introduce the model of disorder, calculate the disorder-averaged Green's functions, the renormalized edge velocity and, finally, 
the edge conductance from Kubo formula. Section \ref{Results} summarizes our results.

\section{Edge states in insulating single-layer graphene}
\label{Edge}

\subsection{Boundary problem}
We begin by analyzing the edge states in disorder-free graphene described by the effective four-band Hamiltonian: 
\begin{equation}
\hat{H} = v_0 \tau_z \, \mbox{\boldmath$\sigma$}{\bf p} + \Delta\tau_z \, \sigma_z, \qquad {\bf p}=(-i\hbar\partial_x, -i\hbar\partial_y, 0),
\label{H}
\end{equation}
where Pauli matrices $\tau_z$ and $\sigma_{x,y,z}$ represent the valley and sublattice degrees of freedom, respectively 
[throughout the paper products of $\tau$ - and  $\sigma$- matrices should be understood as direct products],   
$v_0$ is the bulk Fermi velocity determined by the nearest-neighbor-hopping energy and lattice
constant, and $\Delta$ is the staggered (e.g. substrate-induced \cite{BN_Gio07}) sublattice potential. 
Equation (\ref{H}) adopts the following convention for the basis states:
\begin{equation}
\left(
\begin{array}{c}
\psi_{A_+} \\
\psi_{B_+} \\
\psi_{B_-} \\
\psi_{A_-} 
\end{array}
\right),
\label{Psi}
\end{equation}
where $A,B$ and $\pm$ label the sublattices and valleys, respectively.

Following our previous work on the edge states in semi-metallic graphene~\cite{GT07,GT09a,GT09b,GT09c} 
we will work with the Green's function, ${\hat g}({\bf r},{\bf r}^\prime)$, defined by the equation  
\begin{eqnarray}
[\epsilon\, \hat{I}  - \hat{H}]{\hat g}({\bf r},{\bf r}^\prime)= \hat{I} \delta( {\bf r}-{\bf r}^\prime ),
\label{g_eq}
\end{eqnarray}
where energy $\epsilon$ includes an infinitesimal imaginary part $i\delta$, with $\delta>0$ ($\delta<0$) for the retarded (advanced) Green's function, 
and $\hat{I}=\tau_0\sigma_0$ is a unit matrix composed of the unit matrices in valley ($\tau_0$) and sublattice ($\sigma_0$) spaces.
Equation (\ref{g_eq}) will be solved in a semispace $-\infty < x < \infty, 0\leq y < \infty$ with a single edge at $y=0$ 
described by the boundary condition~\cite{Akhmerov08,McCann04} (see also Appendix \ref{A_BC1}): 
\begin{eqnarray}
&&
g|_{y=0}=- 
\left[
\frac{\tau_0 + \tau_z}{2}\mbox{\boldmath$\sigma$}{\bf n}_+  
+  
\frac{\tau_0 - \tau_z}{2}\mbox{\boldmath$\sigma$}{\bf n}_- 	        
\right]
g|_{y=0}.
\label{BC}\\
&&
{\bf n}_\pm =(n_x,0, \mp n_z),\quad  {\bf n}^2_\pm = n^2_x + n^2_z = 1.
\label{n}
\end{eqnarray}
It involves two unit vectors ${\bf n}_\pm$, orthogonal to each other and to the vector normal to the boundary ${\bf n}_B \, || \, y$, 
ensuring the vanishing of the particle current normal to the edge~\cite{Akhmerov08,McCann04}.
The vector components $n_x$ or $n_z$ serve to parametrize the boundary types considered below in Sec. \ref{ZZ} and \ref{MC}.  

The only restrictions on the boundary condition (\ref{BC}) are the time-reversal symmetry and the absence of the intervalley coupling.  
Thus, Eq. ~(\ref{BC}) can be seen as a generalized continuum model for zigzag-type edges. It cannot be applied 
to the armchair edges because the latter couples the $K$ and $K^\prime$ valleys. 
However, extended armchair edges are unlikely to occur because they are less stable  
than zigzag edges (see, e.g., Ref.~\onlinecite{Girit09}). As to the short-length armchair edges, they can be treated as a special type of the boundary defects 
causing inter-valley scattering, which is considered later in Sec.~\ref{Conductance}.
Therefore, Eq.~(\ref{BC}) is a good starting point for analyzing defect-free graphene. 
It is also easy to verify that Eq.~(\ref{BC}) ensures vanishing of the normal component of the particle current $j_y(x,0)$.

\subsection{Green's function solution} 
  
The Green's function is block-diagonal in valley space,
\begin{eqnarray}
\hat{g}=
\left(
\begin{array}{cc}
\hat{g}^+ & 0\\
  0 & \hat{g}^-
\end{array}
\right),
\label{g_diag}
\end{eqnarray}
where $\hat{g}^\pm$ are $2\times 2$ matrices which, in the basis defined by Eq.~(\ref{Psi}), have the following structures:   
\begin{eqnarray}
\hat{g}^+=
\left(
\begin{array}{cc}
g^+_{AA} & g^+_{AB}\\
g^+_{BA} & g^+_{BB}
\end{array}
\right), 
\qquad 
\hat{g}^-=
\left(
\begin{array}{cc}
g^-_{BB} & g^-_{BA}\\
g^-_{AB} & g^-_{AA}
\end{array}
\right). 
\label{g^pm}
\end{eqnarray}
Let us first calculate the matrix elements of $\hat{g}^+$. Expanding
\begin{eqnarray*}
\hat{g}^+({\bf r}, {\bf r}^\prime)=\sum_k \hat{g}^+_k(y,y^\prime)\,e^{ik(x-x^\prime)}/L,
\end{eqnarray*}
where $L$ is the edge length, and writing Eq.~(\ref{g_eq}) in components, 
it is straightforward to express the off-diagonal elements $g^+_{AB|k}$ and $g^+_{BA|k}$
in terms of the diagonal ones as follows
\begin{eqnarray}
\hat{g}^+_k=\left(
\begin{array}{cc}
\, \hskip 0.7cm g^+_{AA|k} & \frac{v_0 p_-}{\epsilon - \Delta}\, g^+_{BB|k}\\
\frac{v_0 p_+}{\epsilon + \Delta}\, g^+_{AA|k} & \, \hskip 0.7cm g^+_{BB|k}
\end{array}
\right),
\, 
p_\pm = \hbar k \pm \hbar\partial_y,
\label{g^+}
\end{eqnarray}
where $g^+_{AA|k}$ and $g^+_{BB|k}$ satisfy the equations: 
\begin{eqnarray}
&&
[\partial^2_y - q^2 ] g^+_{AA|k}(y,y^\prime)=
\frac{\epsilon + \Delta}{\hbar^2v^2_0} \delta( y-y^\prime ),
\label{g_AA_eq}\\
&&
[\partial^2_y - q^2 ]g^+_{BB|k}(y,y^\prime)=
\frac{\epsilon - \Delta}{\hbar^2v^2_0}\delta( y-y^\prime ),
\label{g_BB_eq}\\
&&
q = \sqrt{ k^2 + \frac{ \Delta^2-\epsilon^2 }{\hbar^2v^2_0} }.
\label{q}
\end{eqnarray}
The boundary conditions for Eqs.~(\ref{g_AA_eq}) and (\ref{g_BB_eq}) follow from Eqs.~(\ref{BC}), (\ref{g_diag}) and (\ref{g^pm}) as
\begin{eqnarray}
&&
\frac{\partial}{\partial y} g^+_{AA|k}(0,y^\prime) =\kappa_A\, g^+_{AA|k}(0,y^\prime),
\label{BC_A}\\
&&
\kappa_A = -\frac{1-n_z}{n_x} \frac{ \epsilon + \Delta}{\hbar v_0} - k,
\label{k_A}\\
%
&&
\frac{\partial}{\partial y} g^+_{BB|k}(0,y^\prime)=\kappa_B\, g^+_{BB|k}(0,y^\prime),
\label{BC_B}\\
&&
\kappa_B = \frac{1+n_z}{n_x} \frac{ \epsilon - \Delta}{\hbar v_0} + k.
\label{k_B}
\end{eqnarray}
We seek the solutions to Eqs.~(\ref{g_AA_eq}) and (\ref{g_BB_eq}) in the form of the linear combinations: 
\begin{eqnarray*}
&&
g^+_{AA|k}(y,y^\prime)=-\frac{\epsilon + \Delta}{2\hbar^2 v^2_0q}{\rm e}^{ -q|y-y^\prime| } + C_A(y^\prime){\rm e}^{-qy}
\\
&&
g^+_{BB|k}(y,y^\prime)=-\frac{\epsilon - \Delta}{2\hbar^2 v^2_0q}{\rm e}^{ -q|y-y^\prime| } + C_B(y^\prime){\rm e}^{-qy}
\end{eqnarray*}
where the first terms are the Green's functions of the unterminated sublattices, while
the second ones are the decaying solutions of the corresponding homogeneous equations.  
The coefficients $C_{A,B}$ are obtained from boundary conditions (\ref{BC_A}) 
and (\ref{BC_B}) with the following results:
\begin{eqnarray}
&&
g^+_{AA|k}(y,y^\prime)=\frac{\epsilon +\Delta}{2\hbar^2v^2_0q}
\left(
{\rm e}^{-q(y+y^\prime)} - {\rm e}^{ -q|y-y^\prime| }
\right)
\nonumber\\
&&
+
\frac{ (1+n_z)(q + k) + n_x (\Delta + \epsilon)/\hbar v_0 }
{ 2( \epsilon - n_z \Delta  + n_x\hbar v_0 k ) }
\,{\rm e}^{-q(y+y^\prime)},
\label{g_AA}
\end{eqnarray}
\begin{eqnarray}
&&
g^+_{BB|k}(y,y^\prime) =\frac{\epsilon -\Delta}{2\hbar^2v^2_0q}
\left(
{\rm e}^{-q(y+y^\prime)} - {\rm e}^{ -q|y-y^\prime| }
\right)
\nonumber\\
&&
+
\frac{ (1-n_z)(q - k) + n_x (\Delta - \epsilon)/\hbar v_0 }
{ 2( \epsilon - n_z \Delta  + n_x\hbar v_0 k ) }
\,{\rm e}^{-q(y+y^\prime)}. 
\label{g_BB}	
\end{eqnarray}
The first terms in Eqs.~(\ref{g_AA}) and (\ref{g_BB}) vanish at the boundary $y=0$ and do not have  poles within the gap, $|\epsilon| < \Delta$, 
implying that the edge states are entirely described by the second terms. 
The latter have a pole within the gap at $\epsilon = n_z \Delta  - n_x\hbar v_0 k$.
Assuming that the energy is close to this pole, we can neglect the first terms in Eqs.~(\ref{g_AA}) and (\ref{g_BB}) and 
find a compact expression for the matrix $\hat{g}^+_k$ [see Eq.~(\ref{g^+})]: 

\begin{equation}
\hat{g}^+_k(y,y^\prime) = (\sigma_0 - \mbox{\boldmath$\sigma$}{\bf n}_+)
\frac{ \Theta(  n_z k + n_x \Delta/\hbar v_0   ) }
     { \epsilon - n_z \Delta  + n_x \hbar v_0 k }
\,q\, {\rm e}^{-q(y+y^\prime)}.
\label{g^+_res}
\end{equation}
The expression for $\hat{g}^-_k$ can be obtained from Eq.~(\ref{g^+_res}) by replacing $v_0 \to -v_0$, $\Delta \to - \Delta$, and $n_z \to -n_z$: 

\begin{equation}
\hat{g}^-_k(y,y^\prime) = (\sigma_0 - \mbox{\boldmath$\sigma$}{\bf n}_-)
\frac{ \Theta( -n_z k + n_x \Delta/\hbar v_0  ) }
     { \epsilon - n_z \Delta  - n_x \hbar v_0 k }
\,q\, {\rm e}^{-q(y+y^\prime)}.
\label{g^-_res}
\end{equation}
The poles in Eqs.~(\ref{g^+_res}) and (\ref{g^-_res}) yield the edge spectrum
in valleys $K$ and $K^\prime$ under the condition that the arguments 
of the Heaviside (theta) functions in the numerators are positive. 
These restrictions are, in turn, enforced by the positiveness of the inverse decay length, $q$.  
Consequently, the edge spectrum is given by

\begin{equation}
\varepsilon_{+,k} = n_z \Delta  - n_x \hbar v_0 k, \,\,\,  n_z k + n_x \frac{\Delta}{\hbar v_0} >0, \,\,\, {\rm for}\,\, {\rm valley}\,\, K, 
\label{E+}
\end{equation}
\begin{equation}
\varepsilon_{-,k} = n_z \Delta  + n_x \hbar v_0 k, \,  -n_z k + n_x \frac{\Delta}{\hbar v_0} >0, \,\,\, {\rm for}\,\, {\rm valley}\,\, K^\prime.
\label{E-}
\end{equation}
These edge states can also be viewed as the solution of a {\em strictly 1D problem} described by Green's functions (\ref{g^+_res}) and (\ref{g^-_res}) 
integrated over the transverse coordinate $y$: 
\begin{equation}
\hat{g}^\pm_k = \int^\infty_0  \hat{g}^\pm_k (y,y)dy = 
\frac{ \sigma_0 - \mbox{\boldmath$\sigma$}{\bf n}_\pm }{ 2 }
\frac{ \Theta(  \pm n_z k + n_x \Delta/\hbar v_0   ) }
     { \epsilon - n_z \Delta  \pm  n_x \hbar v_0 k }.
\label{g^pm_1D}
\end{equation}
Below we analyze these results for two commonly considered confinement types, viz. the zigzag edge and the mass confinement.

\subsection{Zigzag edge}
\label{ZZ}

\begin{figure}[t]
\begin{center}
\includegraphics[width=85mm]{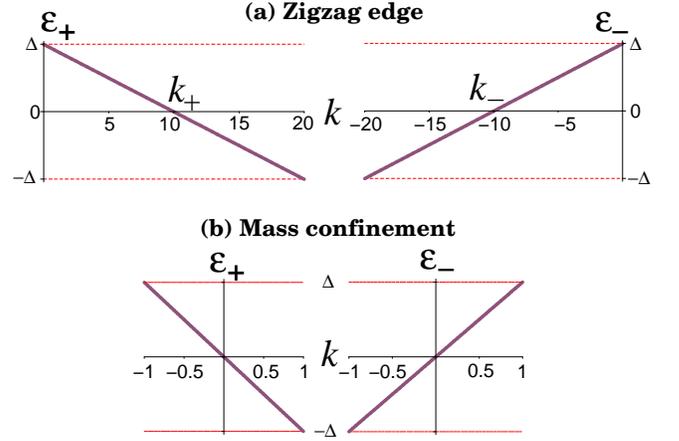}
\end{center}
\caption{
Edge states $\varepsilon_{+}$(k) and $\varepsilon_{-}(k)$ within bulk band gap $[-\Delta,\Delta]$ in valleys $K$ and $K^\prime$, respectively: 
(a) zigzag boundary [see Eqs. (\ref{E_ZZ}) and (\ref{v_ZZ}) with $n_z = 0.99$] and (b) mass confinement [see e.g. Eq. (\ref{E_M})]. 
Edge momentum $k$ is measured in units of $\Delta/\hbar v_0$. 
}
\label{Fig_E}
\end{figure}

In our parametrization of the boundary condition, the zigzag-type edge corresponds to the limit $n_z \to 1$. 
In this case the edge-state spectrum, Eqs. (\ref{E+}) and (\ref{E-}), reduces to 

\begin{eqnarray}
\varepsilon_\pm(k) = \Delta  \mp \hbar v k, \quad  \pm k >0,
\label{E_ZZ}
\end{eqnarray}
where $v$ is the edge-state velocity,
\begin{eqnarray}
v = v_0 n_x = v_0 \sqrt{1-n^2_z}  \ll v_0.
\label{v_ZZ}
\end{eqnarray}
It vanishes for $n_z=1$, which corresponds to the flat edge-state band. In what follows we keep the velocity $v$ finite (but small compared with the bulk velocity $v_0$), 
so that the edge states remain dispersive. We also note that the staggered sublattice potential shifts the edge states by energy $\Delta$ such that 
they cross the mid-gap energy at finite wave-vectors $k_\pm$ [defined in Eq. (\ref{Cond})] with respect to the $K$ and $K^\prime$ points (Fig. \ref{Fig_E}a).
The energy shift reflects the fact that the zigzag edge states reside on one of the sublattices. 
This is seen from the matrix structure of the edge Green's function [Eqs.~(\ref{g_diag}) and (\ref{g^pm_1D})] 
which for $n_z \to 1$ has only two diagonal nonzero matrix elements. 
To simplify the model, from now on we will work with the effective 1D Green's function:~\footnote{
This is justified because of the small edge-state width $\sim \hbar v/\Delta \propto \sqrt{1-n^2_z} \to 0$ as $n_z \to 1$, 
so that it can be made smaller than other length scales in the $y$-direction.  
} 
\begin{eqnarray}
\hat{g}_k = 
\left(
\begin{array}{cccc}
g^+_k & 0 & 0 & 0\\
0 & 0 & 0 & 0\\
0 & 0 & 0 & 0\\
0 & 0 & 0 & g^-_k
\end{array}
\right)=
\hat{P}^+_\tau \hat{P}^+_\sigma g^+_k + \hat{P}^-_\tau \hat{P}^-_\sigma g^-_k,
\label{g_ZZ}
\end{eqnarray}
where $\hat{P}^\pm_\tau$ and $\hat{P}^\pm_\sigma$ are the projector matrices in the valley and sublattice spaces, respectively: 
\begin{equation}
\hat{P}^{\pm}_\tau=\frac{\tau_0 \pm \tau_z}{2}, \quad \hat{P}^{\pm}_\sigma=\frac{\sigma_0 \pm \sigma_z}{2},
\label{P}
\end{equation}
matrix elements $g^+_k$ and $g^-_k$ are given by
\begin{eqnarray}
g^\pm_k=\frac{\Theta(\pm k)}{ \epsilon - \Delta \pm \hbar v k },
\label{g_pm}
\end{eqnarray}
Equation (\ref{g_pm}) describes the left-moving ($+$) and right-moving ($-$) states originating from valleys $K$ and $K^\prime$, respectively. 
Comparing Eqs.~(\ref{g_diag}) and (\ref{g^pm}) with (\ref{g_ZZ}), we see that both left- and right-movers reside on the {\em same} sublattice $A$ 
(the case of sublattice $B$ corresponds to the boundary condition with $n_z\to -1$), as we mentioned in the introduction.

\subsection{Mass confinement}
\label{MC}

This confinement type \cite{Berry87} is realized in the limit $n_z \to 0$, resulting in the edge-state spectrum,
\begin{eqnarray}
\varepsilon_{\pm, k} = \mp \hbar v_0 k. 
\label{E_M}
\end{eqnarray}
These edge states have the velocity equal to the bulk one, $v_0$, and cross the mid-gap energy at points $K$ and $K^\prime$. 
Unlike the zigzag edge, there is no energy shift due to the staggered potential because in this case the edge states propagate on two sublattices. 
This is again seen from the matrix structure of the edge Green's function [Eqs.~(\ref{g_diag}) and (\ref{g^pm_1D})] 
which for $n_z \to 0$ has both diagonal and off-diagonal matrix elements in each valley: 
\begin{eqnarray}
\hat{g}_k &=& 
\left(
\begin{array}{cccc}
g^+_k & -g^+_k & 0 & 0\\
-g^+_k & g^+_k & 0 & 0\\
0 & 0 & g^-_k & -g^-_k\\
0 & 0 & -g^-_k & g^-_k
\end{array}
\right) =
\nonumber\\
&=&
\hat{P}^+_\tau (\sigma_0 - \sigma_x) g^+_k + \hat{P}^-_\tau (\sigma_0 - \sigma_x) g^-_k,
\label{g_M}
\end{eqnarray}
where the valley projectors $\hat{P}^\pm_\tau$ are defined in Eq. (\ref{P}) and matrix elements $g^\pm_k$ are given by
\begin{eqnarray}
&&
g^\pm_k=\frac{1}{\epsilon \pm \hbar v_0 k }.
\label{g^pm_M}
\end{eqnarray}
Equations (\ref{g_M}) and (\ref{g^pm_M}) recover the corresponding results of Ref. \onlinecite{GT10} for the edge states 
in 2D topological insulators. Their transport properties are reviewed in detail in Ref. \onlinecite{GT12}.

\section{Edge conductance}
\label{Conductance}

\begin{figure}[t]
\begin{center}
\includegraphics[width=85mm]{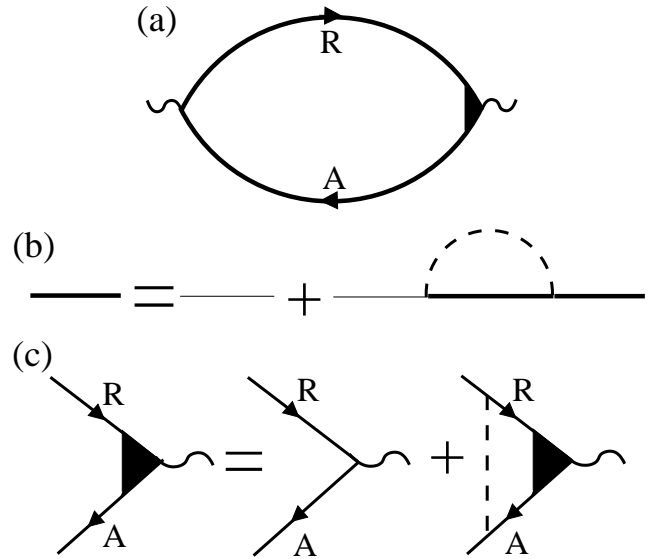}
\end{center}
\caption{
Diagrammatic representations of (a) Kubo formula (\ref{Cond_Kubo}) for the conductance, 
(b) Dyson equation (\ref{G_eq}) for disorder-averaged Green's functions (thick lines) and 
(c) equation (\ref{V_eq}) for renormalized current vertex in ladder approximation. 
Thin and dashed lines correspond to the unperturbed Green's function and disorder correlator, 
respectively.
}
\label{Diagrams}
\end{figure}

In the rest of the paper we focus specifically on transport properties of the edge states on a zigzag-type boundary. 
In the metallic regime the conductance of the edge states is given by Kubo formula: 
\begin{eqnarray}
{\cal G}=\frac{e^2\hbar}{2\pi L}
\int\frac{d k}{2\pi}
{\rm Tr}
\left[
\hat{v}\, \hat{G}^R_k(\epsilon=0) \, \hat{\cal V}_k \, \hat{G}^A_k(\epsilon=0)
\right],
\label{Cond_Kubo}
\end{eqnarray}
where $\hat{G}^{R,A}_k$ are the disorder-averaged retarded and advanced edge Green's functions, 
$\hat{v}$ and $\hat{\cal V}_k$ are the bare and renormalized edge-state velocity operators, and $L$ is 
the length of the edge (e.g. the distance between the source and drain contacts). 
To evaluate Eq. (\ref{Cond_Kubo}) we need to calculate first $\hat{G}^{R,A}_k$ and $\hat{\cal V}_k$ in the presence 
of disorder.

\subsection{Potential and inter-valley coupling disorder}
\label{Disorder}

We assume that edge disorder can be described by equation 
\begin{eqnarray}
\hat{V}(x) = \hat{I} V(x) + \sum_{\mu=x,y}\sum_{\alpha=0,x,y} W_{\mu\alpha}(x) \tau_\mu \sigma_\alpha,
\label{V}
\end{eqnarray}
where the first diagonal term accounts for smooth random potential fluctuations, 
e.g. due to remote ionized impurities in the substrate. 
The potential $V(x)$ is characterized by the correlation function 
\begin{equation}
\langle V(x)V(x^\prime)\rangle = u(x-x^\prime),
\label{VV}
\end{equation}
with averaging $\langle ... \rangle$ over disorder configurations (e.g. over impurity coordinates).
The second term in Eq. (\ref{V}) describes atomically sharp defects (lattice vacancies, short-length armchair edges etc.) which couple the valleys 
(and, generally, the sublattices). For this disorder type we use the correlation function  
\begin{equation} 
\langle W_{\mu\alpha}(x)W_{\mu^\prime\alpha^\prime}(x^\prime) \rangle = 
w(x-x^\prime)\delta_{\mu\mu^\prime}\delta_{\alpha\alpha^\prime},
\label{WW}
\end{equation}
where the Kronecker symbols imply completely uncorrelated valley $\tau_\mu$ and sublattice $\sigma_\alpha$ disorder components. 
The calculations below are done for spatially isotropic disorder with $u(x-x^\prime)=u(x^\prime-x)$ and $w(x-x^\prime)=w(x^\prime-x)$.

\subsection{Disorder-averaged Green's function} 
\label{G_function}
Within the standard self-consistent Born approximation~\footnote{
This method cannot, however, be used for edge states in the strong localization regime studied, e.g. in Ref. \onlinecite{Waka09}. 
}
the disorder-averaged Green's functions  
can be obtained from the Dyson equation (see diagram in Fig. \ref{Diagrams}b) 
which (with suppressed superstripts $R,A$ for brevity) is given by
\begin{equation}
\hat{G}_k=\hat{g}_k + \hat{g}_k\hat{\Sigma}_k\hat{G}_k,
\label{G_eq}
\end{equation}
where $\hat{\Sigma}_k$ is the self-energy:
\begin{eqnarray}
\hat{\Sigma}_k=\int\frac{dk^\prime}{2\pi} 
[u_{k-k^\prime}\hat{G}_{k^\prime} + w_{k-k^\prime}\sum_{\mu\alpha}\tau_\mu\sigma_\alpha \hat{G}_{k^\prime} \tau_\mu\sigma_\alpha ],
\label{Sigma}
\end{eqnarray}
and $u_{k-k^\prime}$ and $ w_{k-k^\prime}$ are the Fourier transforms of the correlation functions $u(x-x^\prime)$ and $w(x-x^\prime)$ 
[see Eqs. (\ref{VV}) and (\ref{WW})]. We seek the solution to Eq. (\ref{G_eq}) in the form of the projector expansion:
\begin{eqnarray}
\hat{G}_k = \hat{P}^+_\tau \hat{P}^+_\sigma G^+_k + \hat{P}^-_\tau \hat{P}^-_\sigma G^-_k, 
\label{G_Ansatz}
\end{eqnarray}
with two unknown scalar functions $G^+_k$ and $G^-_k$. The ansatz (\ref{G_Ansatz}) is valid only for the zigzag-type edge [Cf. Eq. (\ref{g_ZZ})].
Inserting this into Eq. (\ref{Sigma}) we have
\begin{eqnarray}
\hat{\Sigma}_k &=& 
\hat{P}^+_\tau \hat{P}^+_\sigma \int\frac{dk^\prime}{2\pi} [ u_{k-k^\prime} G^+_{k^\prime} + \sum\limits_{\mu,\alpha=1,2} w_{k-k^\prime} G^-_{k^\prime} ] 
\nonumber\\
&
+
&
\hat{P}^-_\tau \hat{P}^-_\sigma \int\frac{dk^\prime}{2\pi} [ u_{k-k^\prime} G^-_{k^\prime} +  \sum\limits_{\mu,\alpha=1,2} w_{k-k^\prime} G^+_{k^\prime}  ]
\nonumber\\
&
+
&
\hat{P}^-_\tau \hat{P}^+_\sigma  \sum\limits_{\mu=1,2} \int\frac{dk^\prime}{2\pi} w_{k-k^\prime} G^+_{k^\prime}  
\nonumber\\
&
+
&
\hat{P}^+_\tau \hat{P}^-_\sigma \sum\limits_{\mu=1,2} \int\frac{dk^\prime}{2\pi} w_{k-k^\prime} G^-_{k^\prime}.  
\label{Sigma_1}
\end{eqnarray}
Here the first two terms include the potential scattering and inter-valley coupling between the left and right movers, $G^+_{k^\prime}$ and $G^-_{k^\prime}$.
This coupling originates from those disorder terms which swap both the valleys and sublattices.  
The other two terms in Eq.~(\ref{Sigma_1}) result from the disorder which swaps either the valleys or the sublattices. 
The latter has no effect on $\hat{G}_k$ because upon inserting Eq.~(\ref{Sigma_1}) into Eq.~(\ref{G_eq}) 
the corresponding products of projectors vanish: $\hat{P}^\pm_\tau \hat{P}^\pm_\sigma \hat{P}^-_\tau \hat{P}^+_\sigma =0$.  
Notice that the summation over the valley and sublattice indices in Eq.~(\ref{Sigma_1}) yields the factor of 4. 

Inserting Eq.~(\ref{Sigma_1}) into Eq.~(\ref{G_eq}) and collecting coefficients at $\hat{P}^\pm_\tau \hat{P}^\pm_\sigma$ 
we obtain algebraic equations for $G^\pm_k$:
\begin{eqnarray}
G^+_k=\frac{1}{ \hbar v (k-k_+) - \Sigma^+_k},\,\,  G^-_k=\frac{1}{ \hbar v (k_- - k) - \Sigma^-_k },
\label{G_pm}
\end{eqnarray}
where $\Sigma^\pm_k$ are scalar functions given by
\begin{equation}
\Sigma^\pm_k=\int\frac{dk^\prime}{2\pi} [ u_{k-k^\prime} G^\pm_{k^\prime} + 4w_{k-k^\prime} G^\mp_{k^\prime} ]. 
\label{Sigma_pm}
\end{equation}
Like in conventional metals (see e.g. Ref. \onlinecite{Rammer}), Eqs. (\ref{G_pm}) can now be solved using the sharpness of the Green's functions $G^\pm_k$ 
near Fermi points $k_\pm$, which yields the finite quasiparticle life-time $\tau$:
\begin{eqnarray}
G^+_k=\frac{1}{ \hbar v (k-k_+) + \frac{s i \hbar}{2\tau} },\,\,  G^-_k=\frac{1}{ \hbar v (k_- - k) + \frac{s i \hbar}{2\tau} },
\label{G_pm_1}
\end{eqnarray}
where the sign $s$ is positive or negative for the retarded or advanced functions, respectively.
The potential and inter-valley scattering mechanisms give additive contributions to the spectral broadening:
\begin{eqnarray}
&&
\frac{1}{\tau} =  \frac{1}{ \tau_{_V} } + \frac{1}{ \tau_{_{KK^\prime}} }, 
\label{Gamma}\\
&&
\frac{1}{ \tau_{_V} }= u_0 \frac{2\pi N}{\hbar}, \quad \frac{1}{\tau_{ _{KK^\prime} } }=4w_{k_+ - k_-} \frac{2\pi N}{\hbar},
\label{tau}
\end{eqnarray}
where $N$ is the density of states per valley and spin.
Since potential disorder cannot cause backscattering,
the corresponding scattering rate $1/\tau_{_V}$ involves the correlation function $u_{k-k^\prime}$ at zero momentum transfer $k-k^\prime=0$, 
which corresponds to forwardscattering. 
In contrast, inter-valley scattering occurs between the Fermi points $k_\pm$ (see Fig. \ref{Fig_E}), 
so that the corresponding scattering rate $1/\tau_{ _{KK^\prime} }$
involves the correlation function $w_{k_+ - k_-}$ with finite momentum transfer $k_+ - k_-= 2 \Delta/\hbar v$.

\subsection{Vertex renormalization}
\label{Vertex}
We demonstrate below the interplay of the scattering times $\tau, \tau_{_V}$ and $\tau_{_{KK^\prime}}$ [Eqs. (\ref{Gamma}) and (\ref{tau})] 
in the disorder-renormalized velocity $\hat{\cal V}_k$, which is one of the central result of this paper. 
In order to calculate the renormalized velocity $\hat{\cal V}_k$ in the conductance formula (\ref{Cond_Kubo}) 
we consider the vertex equation in the usual ladder approximation (see e.g. Ref. \onlinecite{Rammer} and diagram in Fig. \ref{Diagrams}c):
\begin{eqnarray}
\hat{\cal V}_k &=& \hat{v}
+ \int\frac{dk^\prime}{2\pi} 
[u_{k-k^\prime}\hat{G}^R_{k^\prime}\hat{\cal V}_{k^\prime}\hat{G}^A_{k^\prime}
\label{V_eq}\\
&+& w_{k-k^\prime}
\sum\limits_{\mu,\alpha} 
\tau_\mu\sigma_\alpha 
\hat{G}^R_{k^\prime}\hat{\cal V}_{k^\prime}\hat{G}^A_{k^\prime} 
\tau_\mu\sigma_\alpha ],
\nonumber\\
\hat{v} &=& v (\hat{P}^+_\tau \hat{P}^+_\sigma - \hat{P}^-_\tau \hat{P}^-_\sigma).
\label{v}
\end{eqnarray}
Like the edge Green's function (\ref{g_ZZ}) the bare edge velocity matrix $\hat{v}$ (\ref{v}) has only two diagonal elements $v$ and $-v$ 
corresponding to two counter-prapagating channels from different valleys.   
We seek the solution to Eq. (\ref{V_eq}) in the form of the projector expansion:
\begin{equation}
\hat{\cal V}_k = \hat{P}^+_\tau \hat{P}^+_\sigma {\cal V}^+_k  - \hat{P}^-_\tau \hat{P}^-_\sigma {\cal V}^-_k +  
                 \hat{P}^-_\tau \hat{P}^+_\sigma {\cal V}^{(1)}_k - \hat{P}^+_\tau \hat{P}^-_\sigma {\cal V}^{(2)}_k,
\label{V_Ansatz}
\end{equation}
with four unknown scalar functions ${\cal V}^+_k$, ${\cal V}^-_k$, ${\cal V}^{(1)}_k$ and ${\cal V}^{(2)}_k$. 
Inserting Eqs.~(\ref{G_Ansatz}), (\ref{v}) and (\ref{V_Ansatz}) into Eq.~(\ref{V_eq}) and collecting the coefficients 
at the projectors, we find that ${\cal V}^{(1,2)}_k$ can be expressed through ${\cal V}^\pm_k$ by means of
\begin{eqnarray}
{\cal V}^{(1,2)}_k =\int\frac{dk^\prime}{2\pi} 2w_{k-k^\prime}G^{\pm R}_{k^\prime}{\cal V}^\pm_{k^\prime}G^{\pm A}_{k^\prime}, 
\label{V^1,2}
\end{eqnarray}
and ${\cal V}^\pm_k$ satisfy the following equations:
\begin{equation}
{\cal V}^\pm_k = v
+ \int\frac{dk^\prime}{2\pi} 
[ u_{k-k^\prime}G^{\pm R}_{k^\prime}{\cal V}^\pm_{k^\prime}G^{\pm A}_{k^\prime}
- 4w_{k-k^\prime} G^{\mp R}_{k^\prime}{\cal V}^\mp_{k^\prime}G^{\mp A}_{k^\prime}].
\label{V^pm}
\end{equation}
Here again the integral over $k^\prime$ can be calculated using the sharpness of the Green's functions at $k=k_\pm$ [see Eqs.~(\ref{G_pm_1}) - (\ref{Gamma})], 
which yields the following result:
\begin{eqnarray}
{\cal V}^\pm_k = v +
\frac{ u_{k-k_\pm} }{ u_0 + 4w_{k_+-k_-} } {\cal V}^\pm_{k_\pm} 
-
\frac{ 4w_{k-k_\mp} }{ u_0 + 4w_{k_+-k_-} } {\cal V}^\mp_{k_\mp}. 
\label{V^pm_1}
\end{eqnarray}
In parallel, we perform the $k$-integration in the conductance formula (\ref{Cond_Kubo}), obtaining
\begin{eqnarray}
{\cal G}=\frac{e^2}{hL} ({\cal V}^+_{k_+} + {\cal V}^-_{k_-})\tau. 
\label{Cond1}
\end{eqnarray}
The required sum of the renormalized velocities at Fermi points $k=k_\pm$ is obtained from Eq. (\ref{V^pm_1}) as 
\begin{eqnarray}
{\cal V}^+_{k_+} + {\cal V}^-_{k_-} = v \,
\frac{ u_0 + 4w_{k_+ - k_-} }{ 4w_{k_+-k_-} } =v\, \frac{ \tau_{ _{KK^\prime} } }{\tau},
\label{V^pm_2}
\end{eqnarray}
yielding, finally, the edge conductance:
\begin{eqnarray}
{\cal G} = \frac{e^2}{h} \frac{ \tau_{  _{KK^\prime} } }{\tau} \frac{v\tau}{L} = 
\frac{e^2}{h} \frac{v\tau_{ _{KK^\prime} } }{L}.
\label{Cond2}
\end{eqnarray}

\section{Results and conclusions}
\label{Results}

We have demonstrated that in single-layer graphene with insulating bulk zigzag-like edges provide pathways for  
metallic conduction. Both intra- and inter-valley scattering have been taken into account.      
Although both scattering mechanisms contribute to the elastic broadening of the spectrum (\ref{Gamma}),  
only the intervalley scattering time $\tau_{ _{KK^\prime} }$ (\ref{tau}) enters the edge conductance (\ref{Cond2}).
The transport mean-free path can then be identified as $\ell_{ _{KK^\prime} }= v \tau_{ _{KK^\prime} }$.
We emphasize that Eq. (\ref{Cond1}) holds under weak scattering condition 
$|k_\pm|\ell_{_{KK^\prime} } = \Delta \, \tau_{ _{KK^\prime} }/\hbar \gg 1$, 
enforced by the staggered potential $\Delta$ in zigzag-type terminated graphene. 
Let us discuss qualitatively the dependence of the edge conductance on the staggered potential $\Delta$ and disorder strength, 
assuming a Gaussian correlation function for the inter-valley disorder, 
\begin{equation}
w(k)=\sqrt{2\pi} W^2 L_c \exp(-k^2L^2_c/2), 
\label{w_Gauss}
\end{equation}
where $L_c$ is the disorder correlation length and $W$ is the root-mean-square amplitude of the disorder.
From Eqs. (\ref{Cond}) and (\ref{w_Gauss}) we have
\begin{eqnarray}
{\cal G}(\Delta)=\frac{e^2}{h} \frac{\hbar^2 v^2 }{ 4\sqrt{2\pi} L L_c W^2} \exp\left( \frac{ 2 \Delta^2 L^2_c}{ \hbar^2 v^2 } \right). 
\label{Cond_Gauss}
\end{eqnarray}
We see that the edge conductance exponentially increases with $\Delta$. The reason is that the inter-valley scattering involves the finite momentum 
transfer $k_+ - k_- = 2\Delta/\hbar v$ between the Fermi points $k_\pm$, with the scattering probability 
$\propto w(k_+ - k_-)=w(2\Delta/\hbar v)$ reducing with $\Delta$. On the other hand, ${\cal G}$ gets suppressed algebraically as $1/W^2$ 
with increasing root-mean-square disorder amplitude $W$, which may help in practice to reduce the edge conductance.

\acknowledgments
G.T. acknowledges the hospitality of the Max Planck Institute for the Physics of Complex Systems in Dresden where this work was initiated.
M.H. thanks the DFG for support in the Emmy Noether Programme and through Forscherguppe 760.

\appendix

\section{ Derivation of boundary condition (\ref{BC}) }
\label{A_BC1}

We begin by deriving the boundary condition (\ref{BC}) for the Green's function at the sample edge $y=0$. 
To be concrete we consider the retarded Green's function in real space and time $g({\bf r}t,{\bf r}^\prime t^\prime)$. 
It is a $4 \times 4$ matrix in valley and sublattice space, with matrix elements $g_{jj^\prime}({\bf r}t,{\bf r}^\prime t^\prime)$ standardly expressed 
through the annihilation $\Psi_j({\bf r}t)$ and creation $\Psi^\dagger_{j^\prime}({\bf r}^\prime t^\prime)$ field operators as 
\begin{eqnarray}
&&
g_{jj^\prime}({\bf r}t,{\bf r}^\prime t^\prime) = \frac{ \Theta(t-t^\prime) }{ i\hbar } \times
\nonumber\\
&&\times
\langle\langle  
\Psi_j({\bf r}t)\Psi^\dagger_{j^\prime}({\bf r}^\prime t^\prime) + \Psi^\dagger_{j^\prime}({\bf r}^\prime t^\prime)\Psi_j({\bf r}t)
\rangle\rangle,
\label{g_jj}
\end{eqnarray}
where $j$ (and independently $j^\prime$) runs over the index set $A_+, A_-, B_-$ and $B_+$ of the basis states introduced in Eq. (\ref{Psi}), 
the double brackets denote averaging with the equilibrium statistical operator, and $\Theta(t)$ is the Heaviside function. 

It is obvious from Eq. (\ref{g_jj}) that the boundary condition for the Green's function is just the same as for $\Psi_j({\bf r}t)$. 
Indeed, the boundary condition for $\Psi_j({\bf r}t)$ (derived earlier in Refs. \onlinecite{Akhmerov08, McCann04}) can be written as 
\begin{equation}
\Psi_j({\bf r}t)|_{y=0} = M_{ji} \Psi_i({\bf r}t)|_{y=0},
 \label{BC_Psi}
\end{equation}
where $M_{ji}$ are the elements of the $4\times 4$ matrix 
\begin{equation}
M = -\frac{\tau_0 + \tau_z}{2}\mbox{\boldmath$\sigma$}{\bf n}_+  
-  
\frac{\tau_0 - \tau_z}{2}\mbox{\boldmath$\sigma$}{\bf n}_-, 
 \label{M}
\end{equation}
which is defined in Eq. (\ref{BC}) and main text. 
Inserting Eq. (\ref{BC_Psi}) into Eq. (\ref{g_jj}) at $y=0$, 
we obtain the boundary condition for the matrix elements of the Green's function:
\begin{equation}
g_{jj^\prime}({\bf r}t,{\bf r}^\prime t^\prime)_{y=0} = M_{ji} g_{i j^\prime}({\bf r}t,{\bf r}^\prime t^\prime)|_{y=0},
 \label{BC_g_jj}
\end{equation}
which along with Eq. (\ref{M}) yields Eq. (\ref{BC}).

\section{Derivation of boundary conditions (\ref{BC_A}) and (\ref{BC_B}) }
\label{A_BC2}

In order to derive these equations we start with the boundary condition for the upper block 
of the Green's function, $g^+|_{y=0} = -\mbox{\boldmath$\sigma$}{\bf n}_+ g^+|_{y=0}$  [see Eqs. (\ref{BC}), (\ref{g_diag}) and (\ref{g^pm})], 
which has explicit form
\begin{eqnarray}
 \left(
\begin{array}{cc}
g^+_{AA} & g^+_{AB}\\
g^+_{BA} & g^+_{BB}
\end{array}
\right)_{y=0}
=
\left(
\begin{array}{cc}
n_z & -n_x\\
-n_x & -n_z
\end{array}
\right)
\left(
\begin{array}{cc}
g^+_{AA} & g^+_{AB}\\
g^+_{BA} & g^+_{BB}
\end{array}
\right)_{y=0}.
\label{BC_g^+}
\end{eqnarray}
Therefore, for the upper diagonal element we have 
\begin{equation}
g^+_{AA} |_{y=0} = n_z\, g^+_{AA} |_{y=0} - n_x\, g^+_{BA} |_{y=0}.
 \label{BC_AA}
\end{equation}
Expanding in plane waves $e^{ik(x-x^\prime)}$ and using the relation [see Eq. (\ref{g^+})],
\begin{equation} 
g^+_{BA|k}(y,y^\prime)= \frac{\hbar v_0 }{\epsilon + \Delta}\, ( k + \partial_y )g^+_{AA|k}(y,y^\prime), 
\label{g_BA}
\end{equation}
we obtain from Eqs. (\ref{BC_AA}) and (\ref{g_BA}) a closed boundary condition for $g^+_{AA|k}|_{y=0}$, which after elementary algebra yields 
the boundary condition (\ref{BC_A}). Repeating step by step the same calculation for $g^+_{BB} |_{y=0}$ in Eq. (\ref{BC_g^+}) 
leads to the boundary condition (\ref{BC_B}) for the other diagonal matrix element of the Green's function.

\end{document}